\documentclass[10pt,a4paper,aps,showpacs,
							amssymb,pra,floatfix,reprint,
							showkeys,superscriptaddress,
							longbibliography
							]{revtex4-1}

\usepackage[english]{babel}
\usepackage[tbtags,intlimits]{amsmath}

\usepackage[outercaption]{sidecap}

\usepackage{slashed}
\usepackage[usenames,dvipsnames]{xcolor}
\usepackage{hyperref}
\usepackage[squaren]{SIunits}
\usepackage{booktabs}

\usepackage[pdftex]{graphicx}
\usepackage{grffile} 

\usepackage{bbold}

\let\vec\boldsymbol
\let\mathbf\boldsymbol

\newcommand{\op}[1]{\hat{#1}}
\newcommand{\vop}[1]{\vec{\hat{#1}}}
\newcommand{\scp}[2]{\langle #1 | #2 \rangle }

\renewcommand{\d}{  {\mathrm d}   }
\newcommand{\xuvpol}{\vec \varepsilon_{\vec k \Lambda_X}}
\newcommand{\transamp}{\mathcal T}
\newcommand{\twistamp}{a_{\varkappa m}(\vec k_\perp)}
\newcommand{\twistint}{\frac{\mathrm d^2 \vec k_\perp}{(2\pi)^2}}
\newcommand{\dichroism}[1]{\mathcal D^{(#1)}}

\def\bram#1{  \left\langle  #1   \right\vert   }

\def\mem#1#2#3{  \left\langle #1 \left\vert  #2 \right\vert #3 \right\rangle   }

\begin{document}

\title{Two-color above threshold ionization of atoms and ions in XUV Bessel beams {and combined  
       with} intense laser light}

\author{D. Seipt}
\email{d.seipt@gsi.de}
\affiliation{Helmholtz-Institut Jena, Fr{\"o}belstieg 3, 07743 Jena, Germany}
\affiliation{Friedrich-Schiller-Universit\"at Jena, Theoretisch-Physikalisches Institut, 07743 Jena, Germany}

\author{R. A. M\"uller}
\affiliation{Physikalisch--Technische Bundesanstalt, D--38116 Braunschweig, Germany}
\affiliation{Technische Universit\"at Braunschweig, D--38106 Braunschweig, Germany}

\author{A. Surzhykov}
\affiliation{Physikalisch--Technische Bundesanstalt, D--38116 Braunschweig, Germany}
\affiliation{Technische Universit\"at Braunschweig, D--38106 Braunschweig, Germany}

\author{S. Fritzsche}
\affiliation{Helmholtz-Institut Jena, Fr{\"o}belstieg 3, 07743 Jena, Germany}
\affiliation{Friedrich-Schiller-Universit\"at Jena, Theoretisch-Physikalisches Institut, 07743 Jena, Germany}

%
%

%
%

%
%

%
%
\begin{abstract}
The two-color above-threshold {ionization} (ATI) of atoms and ions is investigated for a vortex 
{Bessel beam in the presence of} a strong near-infrared (NIR) light field. While the photoionization is caused 
by the photons from the {weak but extreme} ultra-violet (XUV) vortex Bessel beam, the energy {and angular distribution of}
the photoelectrons {and their sideband structure} are affected by the plane-wave NIR field. We here explore 
the energy spectra and angular emission of the photoelectrons {in such two-color fields as a function}
of the size and location of the target atoms with regard to the beam axis. {In addition,} analogue to the 
circular dichroism in typical two-color ATI {experiments} with circularly polarized {light, we define and
discuss seven} different dichroism signals for such {vortex Bessel beams that arise from} the various combinations 
of the orbital and spin angular momenta of the two light fields. {For localized targets, it}  is found that 
these dichroism signals {strongly depend on} the size and position {of the atoms} relative to the beam. 
For macroscopically extended targets, in contrast, three of these {dichroism} signals tend to zero, while the 
other four {just} coincide with the standard circular dichroism, {similar as for} Bessel beams with small 
opening angle. {Detailed computations of the dichroism are performed} and discussed for the $4s$ valence-shell 
photoionization of Ca$^+$ ions.
\end{abstract}

\maketitle

%
%
%
%
%
%

\section{Introduction}

{Studies on non-perturbative multiphoton processes in intense laser pulses have rapidly advanced during recent
years and helped to explore the inner-atomic motion of electrons at femto- and attosecond timescales 
\cite{Corkum:NaturePhys2007,Krausz:RevModPhys2009,Pazourek:RevModPhys2015}. 
For example, such multiphoton ionization and inner-shell processes have not only been 
observed for noble gases \cite{Gilbertson:PRL2010,Feist:PRL2009}
but also for molecules \cite{Haessler:NaturePhys2010},
surfaces and elsewhere  \cite{Cavalieri:Nature2007,Kruger:JPB2012}. 
Today, these studies enable one to generate quite routinely attosecond pulses by high-order harmonic generation \cite{Sansone:Science2006,Popmintchev:Science2012,Calegari:JPB2016},
or to control the above-threshold ionization (ATI) \cite{Milosevic:JPB2006,Wittmann:NaturePhys2009}. }

In typical ATI experiments, an electron is released from an atom or molecule by absorbing one or several photons from 
a near-infrared (NIR) laser field \textit{more} than required energetically in order to overcome the ionization threshold. 
The ATI energy spectra of the photoelectrons therefore exhibit a series of peaks, just separated by the NIR photon 
energy, while the relative strength of these peaks may depend significantly on the intensity and temporal structure of the 
incident laser pulses. These ATI spectra are thus quite in contrast to the photoelectron spectra as obtained by just weak high-frequency (XUV) radiation, where the absorption of a single photon is sufficient to eject a bound electron and 
where the single photoline (for each possible final state of the photoion) is usually well described by perturbation theory. 
In two-color ATI, such a XUV field is often combined with intense NIR laser pulses in order to investigate the ionization 
of sub-valence electrons: While, under these circumstances, the NIR field alone is not sufficient to ionize the atoms 
or molecules efficiently, it is intense enough to stimulate the absorption or emission of one or several additional 
NIR photons by the outgoing electron. This \textit{non-perturbative} interaction of the electrons with both, the XUV 
\textit{and} NIR fields then leads to the well-known 'sidebands' that typically occur as satellites to the normal 
photolines \cite{Ehlotzky:PhysRep2001,Radcliffe:NJP2012}. Such sidebands were first observed in the 
two-color ATI of noble gases as well as in laser-assisted Auger processes 
\cite{Muller:JPB1986,Schins:PRL1994,Glover:PRL1996,Meyer:PRA2006,Meyer:PRL2008,Meyer:PRL2010}. More recently, these sidebands have 
been applied in pump-probe photoelectron spectroscopy \cite{Helml:NatPhoton2014} or in imaging the molecular orbitals of $\rm H_2O$, $\rm O_2$ and $\rm N_2$ \cite{Leitner:PRA2015}.

{Apart from the temporal structure of the XUV and NIR pulses, the intensity of the sidebands depends of course also
on the relative orientation and linear polarization of these fields \cite{OKeeffe:PRA2004,Guyetand:JPB2005}. This orientation 
dependence has been explored especially by Meyer and coworkers \cite{Meyer:PRL2008} for the angular distribution of the
sidebands in the photoionization of helium. Later, it was shown theoretically \cite{Kazansky:PRA2012} that the 
two-color ATI sideband spectra are rather sensitive also with regard to the circular polarization of both, the XUV and NIR 
light, and an \textit{asymmetry} in the photoelectron spectra was found, if the circular polarization of one of the 
field is changed from same to the opposite direction, a phenomenon that is termed today as circular dichroism in 
two-color ATI. This circular dichroism, that is associated with some flip of the spin-angular momentum (SAM) of the 
incident light field, has recently been utilized, e.g., for \textit{measuring} the polarization state of 
an ultra-violet free electron laser \cite{Kazansky:PRL2011,Kazansky:PRA2012,Mazza:NatCommun2014}. For molecules, 
in addition, the question has been raised how two-color ATI spectra are affected by the molecular symmetry and 
the polarization of the incident radiation \cite{Guyetand:JPB2008,Lux:AngewChemie2012}.}

In this work, we investigate the two-color ATI process if the usual plane-wave XUV field is \textit{replaced} by a 
vortex (Bessel) beam, {also} known as 'twisted light', that carries not only SAM but also orbital angular momentum 
(OAM). Indeed, the study of such OAM light has attracted much recent interest
{for the manipulation of microparticles \cite{Molina:NatPhys2007},
for investigating fundamental interactions \cite{Stock:PRA2015,Quinteiro:PRA2015,Surzhykov:PRA2015,Schmiegelow:NatureCommun2016,Radwell:PRL2015},
and for multiplexing in telecommunication \cite{Bozinovic:Science2013,Krenn:NJP2014}, to name a few.}
In particular, we here explore the energy spectra and angular emission of photoelectrons 
ejected by a vortex Bessel beam in the presence of strong NIR light,
and how these photoelectron spectra depend on the size  
and location of the target (atoms) with regard to the axis of the vortex beam. We assume an XUV Bessel beam that is 
energetic enough to photoionize the atom, while the plane-wave NIR field affects the sideband structure as well as  
the energy and angular distribution of the photoelectrons. Analogously to the circular dichroism from above, we define 
and discuss moreover \textit{seven} possible dichroism signals which arise from different combinations of the orbital 
and spin angular momenta of the two light fields involved. For localized targets, we find that these
dichroism signals sensitively depend on the size and position of the atoms relative to the beam axis. For macroscopically 
extended targets, in contrast, three of these signals tend to zero, while the other four just approach
the (standard) circular dichroism. To discuss these finding, detailed computations of the dichroisms are performed 
and discussed for the $4s$ valence-shell photoionization of Ca$^+$ ions.

{In the next section, we shall first evaluate the transition amplitude and photoionization probability for the 
two-color ATI within the strong-field approximation (SFA), if a vortex Bessel beam interacts with an atom in the 
presence of a strong, plane-wave NIR field. Note that atomic units are used throughout the paper.
Details about the twisted XUV Bessel beam are given in Subsection 
\ref{sect:bessel}, while the choice of targets is explained in {Subsection~\ref{sect:target}.
The} possible dichroism signals for such a two-color field are defined in Section \ref{sect:dichroism}~. Emphasize is placed here especially
on the influence of a \textit{localized \textit{versus} macroscopically extended} target in exploring the sensitivity 
of the dichroism with regard to the target size. Detailed calculations of the photoelectron spectra and the various 
dichroism signals are presented and discussed later in Section~\ref{sect:results}. Finally, conclusions are 
given in Section~\ref{sect:summary}.}

\begin{figure}
\includegraphics[width=0.9\columnwidth]{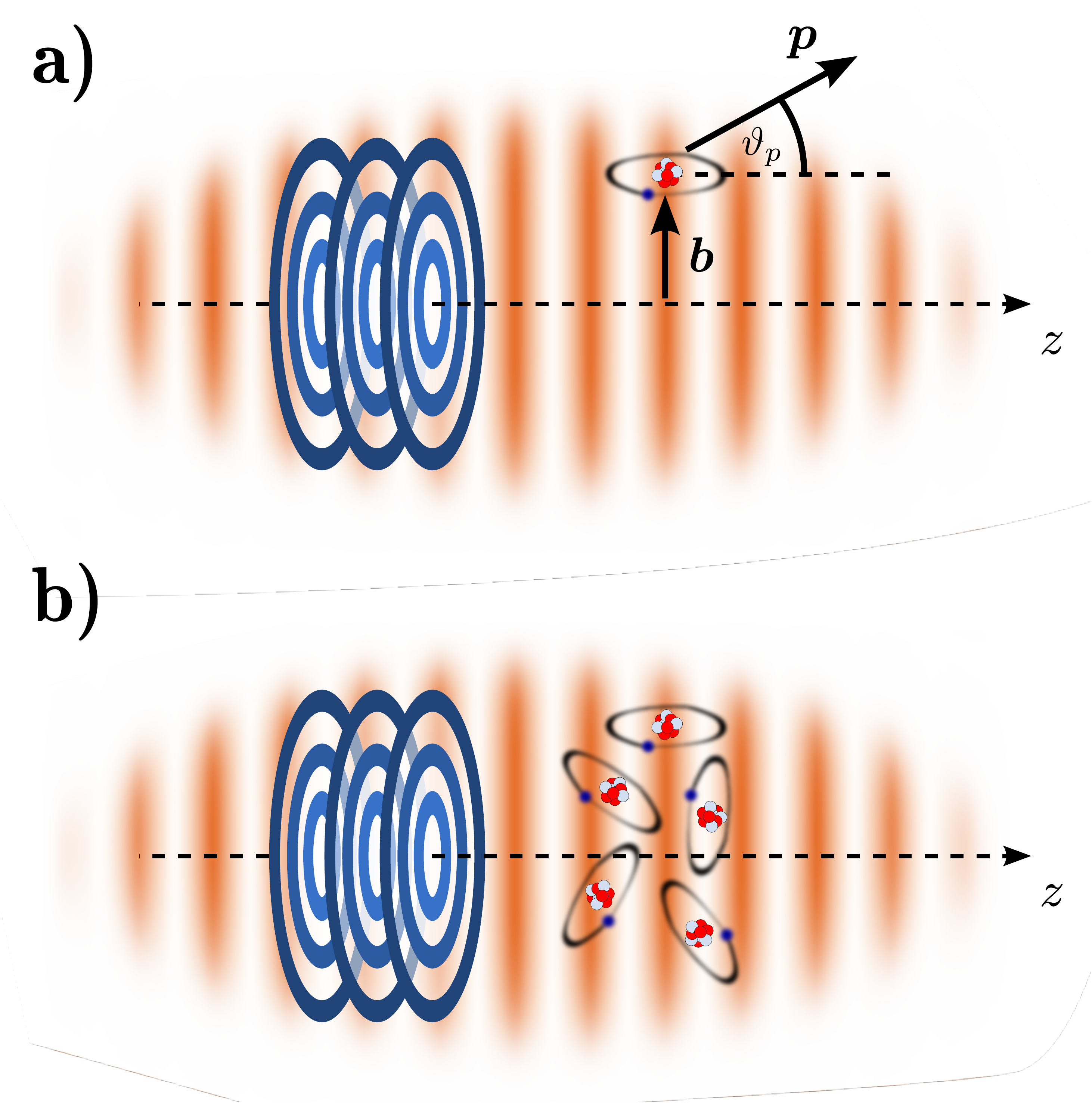}
\caption{{Scheme for} the two-color ATI of an atom by an XUV vortex beam (blue) and in the presence of a strong 
NIR field (orange). The atoms are assumed to be {either} a) localized with regard to the common beam axis 
or b) equally distributed over the cross section of the XUV Bessel beam.}
\label{fig:1}
\end{figure}

\section{Theoretical Background}
\label{sect:theory}

\subsection{Two-color ATI in Strong-Field Approximation}
\label{subsect:system}

We here explore the two-color ATI of atoms (and ions) if they interact with a weak XUV vortex beam and an intense 
plane-wave NIR field. In particular, we assume an (almost) monochromatic XUV vortex beam, as for instance generated 
by free-electron lasers (FEL), and which is energetic enough to ionize the target atom. Although quite strong,
moreover, we suppose some NIR laser pulse with many optical cycles so that it can be described as a monochromatic 
plane-wave. Together, these two assumptions ensure that the 'sideband regime' holds \cite{Radcliffe:NJP2012},
in which photoelectrons are expected not only at the given photoline but also at energies that differ by one or several
energy quanta of the the NIR field. Moreover, both fields are supposed to propagate along a common beam axis that is
taken also as the quantization axis ($z$-axis). Finally, the atomic target is either a microscopic target of trapped atoms or 
ions, localized at some position $\vec b$ in the $xy$-plane perpendicular to the beam axis, 
cf.~Ref.~\cite{Schmiegelow:NatureCommun2016} and Fig.~\ref{fig:1}a, or as \textit{macroscopic} and uniformly distributed 
target over the cross section of the twisted XUV beam (Fig.~\ref{fig:1}b).

{After their interaction with the two-color field, the photoelectrons leave the interaction region with the asymptotic 
(canonical) momentum $\vec p$ as measured at the detectors. Below, we shall analyze the angular and momentum distribution 
of these electrons as a function of the polar angle $\vartheta_p$ of the momentum, i.e.~with regard to the common beams 
axis, as well as for different azimuth angles $\varphi_p$ 
[as defined by the impact vector $\vec b \,=\, (b, \varphi_p = 0, z=0)$ of the target atom]. 
{In particular, we aim} to understand how the photoelectron distributions are affected by the OAM of the 
XUV beam and by relative changes in either the SAM and/or OAM of the two-color fields.}

Within the SFA, the transition amplitude $\transamp$ for the two-color ATI of a single \textit{active} electron, 
being initially in the bound state $ | \phi_0 \rangle$, reads as

\begin{small}
\begin{align}  \label{eq:def:amplitude}
   \transamp (\vec p)  &=  -i \int_{-\infty}^\infty \! \d t \,
			   \mem{ \Psi^{(V)}_{\vec q(t)}  }
			   { \, \vop p \cdot \vec A_X(\vec r) \,  \vphantom{\Psi^{(V)}_{\vec q(t)}  }   }
			   { \phi_0 } \;
		           e^{i\, (E_B-\omega_X) t} \,,
\end{align}
\end{small}

\noindent
where $E_B$ is the binding energy of the active electron and $\vec A_X(\vec r)$ the vector potential of the XUV Bessel
beam with frequency $\omega_X$. We shall describe the details of this vector potential in the next subsection. 
In the SFA, moreover, the final (continuum) state of the electron is typically described by a Volkov state 
$\bram{ \Psi^{(V)}_{\vec q(t)} } = \bram{ \vec q(t) }\,  e^{-iS_V(t)}$ (in length gauge), which neglects the effect 
of the parent ion upon the motion of the liberated electron,
and where $\bram{ \vec q(t) }$ describes {the plane-wave electron wave function} with the kinetic momentum $\vec q(t) = \vec p - \vec A_L(t)$. 
In the presence of an external NIR field $\vec A_L (t)$, this kinetic momentum is different from the \textit{conserved} 
canonical momentum $ \vec p $ {which is} measured at the detector, eventually.
Finally, the phase of the Volkov wavefunction is given by 
\cite{Volkov:1935}
\begin{align}
   S_V(t) &=  \int_t^\infty \! \d t' \, \frac{\vec q(t)^2}{2}
	   =  \frac{1}{2} \int_t^\infty \! \d t' \, \left[ \vec p - \vec A_L(t') \right]^2 \,.
\end{align}
While we have employed the length gauge for the strong assisting NIR laser field $\vec A_L$,
we still use the interaction operator with the high-frequency XUV vortex field $\vec A_X$
{in} the velocity gauge form 
because of the complex spatial structure of {$\vec A_X(\vec r)$}.

\subsection{Characterization of twisted Bessel beams}
\label{sect:bessel}

{So-called twisted or {vortex} (light) beams are known to carry, in addition their possible polarization or SAM, 
also an orbital angular momentum (OAM) owing to their helical phase fronts.
Typically, {twisted} beams show a very 
characteristic annular intensity distribution with zero intensity on the beam axis.
This zero intensity line is called the \textit{vortex line}
of the field and it embodies a phase singularity.
 In the XUV frequency region, twisted light beams have been generated recently by means of undulators 
\cite{Hemsing:NatPhys2013,Bahrdt:PRL2013} or by using high-harmonic generation 
\cite{Zuerch:NatPhys2012,Gariepy:PRL2014,Hernandez-Garcia:PRL2013}. Experimentally, such twisted beams can be prepared in 
different modes with regard to the (components of the) angular momenta that are \textit{conserved} for some given beam. 
In the further analysis, we shall assume Bessel beams that are obtained as non-paraxial solutions of the vector 
Helmholtz equation, and which are classified here by the wave vector components $k_z$ and $\varkappa$ 
[with $\varkappa=|\vec k_\perp|$ and $\vec k \:\equiv\: ( \vec k_\perp, k_z)^T \:=\: (\varkappa, \varphi_k, k_z)$], 
the topological charge $m$ as well as the helicity $\Lambda_X$.
}

Here, we shall restrict ourselves to the representation of the vector potential of these Bessel beams in terms of plane 
waves and how we can distinguish between their spin-angular (polarization) and orbital angular momenta. 
Following Refs.~\cite{Jentschura:PRL2011,Ivanov:PRA2011,Matula:JPB2013}, we can write the vector potential
\begin{align} \label{eq:def:vortex}
   \vec A_X(\vec r)  &=  \int \! \twistint \: \twistamp \:
			 e^{i \vec k \cdot \vec r } \, \vec \varepsilon_{\vec k \Lambda_X } \,,
\end{align}
as a superposition of plane waves with wave vectors $\vec k = (\vec k_\perp,k_z)^T$ and the Fourier coefficients
\begin{align} \label{eq:akm}
   \twistamp &= \sqrt{\frac{2\pi}{\varkappa}} \: (-i)^m e^{i m \varphi_k} \: \delta( k_\perp - \varkappa ) \, ,
\end{align}
and where $\vartheta_k = \arctan \varkappa /k_z$ is the so-called cone \textit{opening} angle. Like in the atomic case, 
the (quantum) number $m$ determines the projection of the OAM or the so-called topological charge and $\xuvpol$ 
the polarization (vector) of the plane-wave components. Obviously, the polarization vector must depend explicitly not only
on the helicity $\Lambda_X$ of the plane-wave components, but also on the angles $\vartheta_k$ and $\varphi_k$,
\begin{align}   \label{eq:xuvpol}
   \vec \varepsilon_{\vec k \Lambda_X} 
   &= \frac{e^{i \Lambda_X \varphi_k}}{\sqrt{2}}
      \left(  \begin{matrix}
                 \cos \vartheta_k \cos \varphi_k - i \Lambda_X \sin \varphi_k 	\\
                 \cos \vartheta_k \sin \varphi_k + i \Lambda_X \cos \varphi_k   \\
                 - \sin \vartheta_k
              \end{matrix}
      \right)  \, ,
\end{align}
due to the transversality condition $\vec k \cdot \xuvpol \: = \: 0$.
Indeed, this definition of the polarization vector $\xuvpol$ ensures that, in the  limit of small 
opening angles $\vartheta_k \to 0$, we obtain the usual polarization vectors for circularly polarized plane waves
\begin{align} 	\label{eq:xuvpol.plane}
   \xuvpol \stackrel{\vartheta_k\to0}{\to} \frac{1}{\sqrt2} \: (1,i\Lambda_X,0)^T \,.
\end{align}
It can be shown \cite{Matula:JPB2013} that the polarization vector $\xuvpol$ from Eq.~\eqref{eq:xuvpol} 
is an eigenvector also of the $z$-component $\op J_z \,=\, \op L_z + \op S_z$ of the \textit{total} angular momentum 
operator with eigenvalues $m_J = \Lambda_X$: $\: \op J_z \: \xuvpol \,=\, \Lambda_X \: \xuvpol$.
With this definition of $\xuvpol$, the Bessel beam {\eqref{eq:def:vortex}} is constructed as an eigenfunction of
the \textit{total} angular momentum projection 
$\hat J_z$ with eigenvalue $m_J = m + \Lambda_X$ \footnote{Note the different definition
of the vector $\xuvpol$ in \cite{Matula:JPB2013} which furnishes a different interpretation of the
quantum number $m = m_J$ as the TAM eigenvalue of the Bessel beam,
but without {any} physical consequences.}:
$$\hat J_z \vec A_X (\vec r) = (m + \Lambda_X) \vec A_X(\vec r)\,.$$

For the sake of completeness, let us write here the vector potential of the XUV Bessel beam explicitly
in cylindrical co-ordinates
\begin{multline} \label{eq:bessel}
   \vec A_X(\vec r) = \sum_{m_s=-1,0,1}  \vec \eta_{m_s} \:
                      \sqrt{ \frac{\varkappa}{2\pi} } \: c_{m_s} \: e^{i k_z z} \: i^{\Lambda_X-m_s}             \\[0.1cm]
                      \times \: e^{i(m+\Lambda_X-m_s)\varphi} \: J_{m+\Lambda_X-m_s} (\varkappa r_\perp) \,.
\end{multline}
In this expression, $\vec \eta_0 = (0,0,1)$ and $\vec \eta_{\pm1} = (1,\pm i,0)/\sqrt{2}$ denote the (spherical) 
unit vectors, and the coefficients are $c_0 = - (\sin \vartheta_k)/\sqrt2$
and $c_{\pm1} = (\cos\vartheta_k \pm \Lambda_X )/2$, respectively. As seen from Eq.~\eqref{eq:bessel}, a Bessel beam
consist of three terms with topological charges $m + \Lambda_X$ and $m+\Lambda_X \pm 1$. The relative 
weight of these terms depend on the opening angle $\vartheta_k$, and only {the one} with the topological charge $m$ 
remains \textit{non-zero} for paraxial beams, i.e.~if $\vartheta_k \ll 1$
\footnote{Strictly speaking, the orbital quantum number $m$ here 
  refers to the dominant component of the three possible projections of the orbital angular momentum, and it coincides with the true OAM only in the paraxial limit $\vartheta_k\ll 1 $.}.

\subsection{Transition amplitude for a well-localized atom in a XUV Bessel beam}

We can use the vector potential of the XUV Bessel beam, Eq.~\eqref{eq:def:vortex}, to evaluate the transition 
amplitude \eqref{eq:def:amplitude} for the two-color ATI of atoms and ions and for the emission of photoelectrons
with well-defined \textit{asymptotic} momentum $\vec p$. To do so, we also need to specify the position of the atom 
with regard to the beam axis, i.e.~in terms of an impact parameter vector
$\vec b \:\equiv\: (b, \varphi_b = 0, b_z = 0)$. 
If $\vec r$ denotes the electronic coordinate with respect to the atomic nucleus, that is the center of the atomic 
potential, we have to replace $\vec r \,\to\, \vec b + \vec r$ in the electron-photon interaction operator. We 
therefore see that, for vortex beams, the transition amplitude generally depends on the location of the atom within the beam,
{as indicated by the subscript $\vec b$ in the notation of the transition amplitude:

\begin{align} \label{eq:twisted:amplitude:superposition}
   \transamp_{\vec b}(\vec p) & =  \int \! \twistint \: \twistamp  \: e^{i \vec k \cdot \vec b} \:
                                   \transamp^{\rm pw}({\vec p,\vec k}) \,.
\end{align}
It is readily expressed as a superposition of typical SFA plane-wave amplitudes
}
{
\begin{multline} \label{eq:amplitude:planewave}
\transamp^{\rm pw}(\vec p, \vec k)  =  -i \int \! \d  t \,  \vec q(t) \cdot \xuvpol \, e^{i(E_B-\omega_X) t - i S_V(t)}\\
	       \times \,	   \mem{ \vec q(t) }{ \: e^{i\,\vec k \cdot \vec r} \:}{ \phi_0 }    \, ,
\end{multline}
just weighted by the Fourier coefficients $\twistamp$ of the Bessel beam and the given phase factor 
$e^{i\vec b\cdot \vec k}$.
An analogue superposition of plane-wave amplitudes was found also for the single-photon 
ionization by light from a vortex beam \cite{Surzhykov:PRA2015}, and this remains true
when adding an assisting laser field, 
at least \textit{within} the SFA.
 Let us mention finally that all the time-dependence resides in the plane-wave 
amplitudes $\transamp^{\rm pw}(\vec p,\vec k)$.
}

\subsubsection{Time-dependence of the Volkov phase: Sideband structure}

To obtain and describe the (well-known) sideband structures in the energy spectrum of the emitted photoelectrons in the
two-color ATI, we next need to specify the vector potential of the strong NIR laser pulse that enters the plane-wave
amplitude $\transamp^{\rm pw}(\vec p,\vec k)$. For the strong laser pulse, we here apply the vector potential
\begin{align}
\vec A_L(t)  &=  A_L \left( \begin{matrix}
                               \cos \omega_L t	           \\[0.1cm]
                               \Lambda_L \sin \omega_L t   \\[0.1cm]
                               0
                            \end{matrix}
                     \right) \, ,
\end{align}
of a plane wave with laser frequency $\omega_L$, helicity $\Lambda_L$ and field amplitude $A_L$. In the plane-wave
amplitude of {Eq.~\eqref{eq:amplitude:planewave}},
moreover, we can cast the Volkov phase factor into the form
\begin{align} \label{eq:jacobi:anger}
   e^{-iS_V(t)} 	
   = e^{i ( \frac{p^2}{2} + U_p ) t  } \sum_{\ell={-\infty}}^\infty  
     J_\ell(\alpha_L) e^{-i \ell (\omega_L t - \Lambda_L    \varphi_p)} 
\end{align}
by applying the Jacobi-Anger expansion \cite{book:Watson},
and where $\alpha_L = ( A_L\, p \sin \vartheta_p ) / \omega_L $ just refers to the amplitude of the classical oscillation 
of an electron in the laser field, while $U_p \,=\, A_L^2/2$ is the ponderomotive potential.

With this reformulation of the Volkov phase factor in Eq.~\eqref{eq:jacobi:anger}, the (strong-field) transition 
amplitude $\transamp^{\rm pw}(\mathbf p, \mathbf k)$ therefore becomes
\begin{widetext}
\begin{align} \label{eq:amplitude:jacobi}
   \transamp^{\rm pw}(\mathbf p, \mathbf k) 
   &=  - i  \sum_\ell  J_\ell(\alpha_L) \:  e^{i \ell  \Lambda_L \varphi_p} \,
       \int \! \d t \: e^{i(\frac{p^2}{2}+U_p+E_B-\omega_X - \ell \omega_L) t } \:
       \vec q(t) \cdot \xuvpol \: \mem{\vec q(t) }{\, e^{i\vec k \cdot \vec r}\,}{ \phi_0 } \,		.
\end{align}
To further simplify this amplitude we next have to analyze the scalar product between the kinetic momentum $\vec q(t)$ 
of the electron and the polarization vector of the twisted light $\xuvpol$ in the following subsection.

Before we shall continue, let us note, that the summations in Eqs.~\eqref{eq:jacobi:anger} and 
\eqref{eq:amplitude:jacobi} formally runs from $\ell = - \infty \ldots \infty$. In practice, however, just a finite number 
of sidebands, $\ell_\mathrm{min} \leq \ell \leq \ell_\mathrm{max}$, can be resolved experimentally, while the magnitude of
these sidebands decays exponentially beyond these cut-off values.
These cut-off values can be determined by either a 
saddle point analysis of the Volkov phase \cite{Lewenstein:PRA1994,Zhang:PRA2013,Seipt:PRA2015,Seipt:NJP2016}
or by just making use of the properties of the Bessel functions 
\cite{book:Watson}. From the prior analysis, we have found these cut-off values as
\begin{align} \label{eq:cutoffs}
   \ell_\mathrm{max/min} 
   = \frac{A_L^2\sin^2\vartheta_p}{\omega_L}  
     \pm \sqrt{ \frac{A_L^4 \sin^4 \vartheta_p}{\omega_L^2} 
            + 2 \frac{A_L^2 \sin^2\vartheta_p}{\omega_L} \: (\omega_X- E_B - U_p)} \,,
\end{align}
\end{widetext}
and where the upper/lower sign refer to the max/min values.

\subsubsection{Angular dependence of the photoelectron emission in the transition amplitude}

The angular distribution of the photoelectrons emitted in the two-color ATI process is mainly determined by the 
scalar product $\vec q(t) \cdot \xuvpol$ in the plane-wave amplitudes
{\eqref{eq:amplitude:jacobi}}. This scalar product 
becomes maximum when the \textit{kinetic} momentum of the photelectron $\vec q(t) \,=\, \vec p - \vec A_L(t)$ is, 
at the moment of the ionization, parallel to the polarization vector of the XUV field. We remember that this
kinetic momentum $\vec q(t) $ differs from the conserved \textit{canonical} momentum $\vec p$ 
{as long as the electron is inside the laser pulse.}

\paragraph{Plane Waves:}
\label{subsubsect:theory:plane}

Here, let us first (re-)consider the scalar product  $\vec q(t) \cdot \xuvpol$ for the case of circularly polarized plane waves \cite{Kazansky:PRA2012}. If the XUV pulse propagates for instance along the $z$-direction, $\vec k = k \vec e_z$,
we can apply the {plane-wave limit of the} XUV polarization vector from Eq.~\eqref{eq:xuvpol.plane} and readily obtain for the scalar product
\begin{align} \label{eq:scalarproduct:plane}
   \vec q(t) \cdot \xuvpol &=  \frac{p}{\sqrt 2} \sin \vartheta_p e^{i\Lambda_X \varphi_p} 
		              -\frac{ A_L}{\sqrt 2} e^{i\Lambda_X\Lambda_L \omega_L t} \,.
\end{align}
Moreover, if we combine this expression with Eq.~\eqref{eq:amplitude:jacobi}, the \textit{plane-wave} transition 
amplitude reads	as
\begin{widetext}
\begin{align} \label{eq:amplitude:plane}
   \transamp^{\rm pw} (\vec p, \vec k = k \vec e_z) 
   &=  -i \sum_\ell \mathcal F_\ell(\Lambda_L,\Lambda_X) \, e^{i ( \ell \Lambda_L +\Lambda_X ) \varphi_p } \:
       \int \!\d t \,  e^{i(\frac{p^2}{2} + U_p + E_B-\omega_X - \ell \omega_L) t }
       \mem{ \vec q(t) }{\, e^{i\vec k \cdot \vec r}\,}{ \phi_0 } \,,
\end{align}
where the \textit{sideband} amplitudes
\begin{align} \label{eq:Fl:plane}
   \mathcal F_\ell(\Lambda_L,\Lambda_X) 
   &=  \frac{1}{\sqrt{2}}
       \Big(   J_\ell (\alpha_L) p \sin \vartheta_p - A_L J_{\ell +\Lambda_L \Lambda_X} (\alpha_L) \Big) \,
\end{align}
describe the strength and the angular distribution of the photoelectrons of the $\ell$-th sideband (ATI-peak).
In order to arrive at Eqs.~\eqref{eq:amplitude:plane} and \eqref{eq:Fl:plane}, we have shifted the summation variable $\ell$ 
in the second term of Eq.~\eqref{eq:Fl:plane}.
\end{widetext}

From the sideband amplitude \eqref{eq:Fl:plane}, we immediately find: (i) Only the central photoline ($\ell=0$) occurs 
with the typical $\mathbb P(\vartheta_p) \propto \sin^2 \vartheta_p$ angular dependence if the laser field vanishes, 
i.e.~for $A_L\to 0$ and $\alpha_L\to 0$. Moreover, (ii) the second term of $\mathcal F_\ell$ in Eq.~\eqref{eq:Fl:plane}
contains the product $\Lambda_L \, \Lambda_X$ of the spin angular momenta (helicities) of the XUV and the
NIR laser fields 
in the \textit{order} of the Bessel function $J$.  Therefore, the angular distribution of the photoelectrons 
differ from each other if $\Lambda_X$ and $\Lambda_L$ have either equal or opposite signs. Indeed, it is the sign 
of $\Lambda_L \, \Lambda_X$ that leads to the circular dichroism in the two-color photoionization of atoms by 
plane-wave radiation \cite{Kazansky:PRA2012,Mazza:NatCommun2014}.

\paragraph{XUV Bessel beams:}
 
Of course, the same scalar product in the plane-wave amplitude \eqref{eq:amplitude:jacobi} becomes much more 
complex for a vortex beam \eqref{eq:def:vortex} since it now depends explicitly on the direction of the momentum 
vector $\vec k \:=\: \vec k(\vartheta_k,\varphi_k)$ of the plane-wave components, forming a cone in momentum space. 
Using expression \eqref{eq:xuvpol}, this product can be evaluated as [compare with Eq.~\eqref{eq:scalarproduct:plane}]
\begin{widetext}
\begin{multline} \label{eq:scp:twisted}
   \vec q(t) \cdot  \xuvpol 
   = \frac{p}{\sqrt 2}\:   \left[ \sin \vartheta_p e^{i\Lambda_X \varphi_p} 
                                   \:-\: 2 \sin \vartheta_p \sin^2 \frac{\vartheta_k}{2} \cos (\varphi_p 
				   \:-\: \varphi_k) e^{i \Lambda_X \varphi_k}
	                           \:-\: \cos \vartheta_p \sin \vartheta_k e^{i\Lambda_X \varphi_k}
	                   \right] \\[0.15cm]
   - \frac{A_L}{\sqrt 2}\: \left[ e^{i \Lambda_X \Lambda_L \omega_L t}
		                  \:-\: \sin^2 \frac{\vartheta_k}{2} 
				        \left( e^{i \omega_Lt } \, e^{i(\Lambda_X - \Lambda_L) \varphi_k}
			                     + e^{-i \omega_L t} \, e^{i(\Lambda_X +\Lambda_L) \varphi_k}			
			                \right)	
	                   \right]	\,.
\end{multline}
If we substitute this expression into Eq.~\eqref{eq:amplitude:planewave},
the transition amplitude for the two-color ATI 
of an atom at position $\vec b$ by a XUV Bessel beam can be written as the superposition 
\eqref{eq:twisted:amplitude:superposition} of the plane-wave transition amplitudes
\begin{align} \label{eq:amplitude:pw:2}
   \transamp^{\rm pw} (\vec p, \vec k) 
   &=  - i \sum_\ell  \mathcal F_\ell(\vartheta_k , \varphi_k;\Lambda_X,\Lambda_L) \:
       e^{i(\ell \Lambda_L + \Lambda_X)\varphi_p} \:
       \int \! \d t \:  e^{i(\frac{p^2}{2} + U_p + E_B-\omega_X - \ell \omega_L ) t }
		    \, \mem{ \vec q(t)}{\: e^{i\vec k \cdot \vec r } \:}{\phi_0}  \,,
\end{align}
and with the \textit{modified} sideband amplitudes
\begin{multline}    \label{eq:sideband:tw}
   \mathcal F_\ell( \vartheta_k , \varphi_k ; \Lambda_X, \Lambda_L)		
   =  J_\ell(\alpha_L) \frac{p}{\sqrt2} 
      \left[ \sin \vartheta_p \:-\: 2 \sin \vartheta_p \sin^2 \frac{\vartheta_k}{2} \cos (\varphi_k 
                              \:-\: \varphi_p) e^{i \Lambda_X (\varphi_k - \varphi_p)}
			      \:-\: \cos \vartheta_p \sin \vartheta_k e^{i \Lambda_X ( \varphi_k - \varphi_p)}	
      \right]  \\[0.15cm]
     - \frac{A_L}{\sqrt{2}}
       \left[ J_{\ell + \Lambda_X \Lambda_L}(\alpha_L)	
	      \:-\: \sin^2 \frac{\vartheta_k}{2}
		    \left\{      J_{\ell+1} (\alpha_L) e^{i(\Lambda_X-\Lambda_L)(\varphi_k-\varphi_p)}
			   \:+\: J_{\ell-1}(\alpha_L) e^{i(\Lambda_X+\Lambda_L)(\varphi_k-\varphi_p)}	
		    \right\}
       \right] \, ,
\end{multline}
which now depends on the particular direction of the wave vector $\vec k$.
\end{widetext}
\subsubsection{Analytical time integration of the two-color ATI transition amplitude}

The plane-wave transition amplitude \eqref{eq:amplitude:pw:2} still contains a time integration which cannot be 
performed in general. However, this time integral can be solved analytically if we assume a sufficiently weak 
assisting NIR laser field, $A_L \ll p$, so that 
the kinetic momentum $\vec q(t)$ of the photoelectron can be reasonably
well approximated by the canonical momentum $\vec p$ in the atomic matrix. This then results also in time-independent atomic matrix elements
\cite{Kazansky:PRA2010,Kazansky:PRA2012}. In the dipole approximation, moreover, we can approximate these matrix elements by
\begin{align}
   \mem{\vec q(t) }{\, e^{i \vec k \cdot \vec r} \,}{ \phi_0 } \:\simeq\: \scp{\vec p}{\phi_0} \, ,
   \label{eq:atomic}
\end{align}
which is valid almost everywhere apart from the region close to the vortex line. 
For $\vec b \,=\, 0$, in contrast, the integral over the transverse momentum $\vec k_\perp$ in {\eqref{eq:twisted:amplitude:superposition}} 
vanishes
when the electric-dipole
approximation is employed, and the leading contribution to the twisted-wave amplitude 
$\transamp_{\vec b \simeq 0} (\vec p)$ will then arise from higher-order multipoles 
\cite{Surzhykov:PRA2015,Schmiegelow:NatureCommun2016}.

With these assumptions about the NIR field, we can perform the time integration in the plane-wave amplitude
\eqref{eq:amplitude:pw:2}
\begin{multline} 	\label{eq:time:integral}
   \int \! \d t \, e^{i t ( \frac{p^2}{2} + U_p + E_B - \omega_X  - \ell \omega_L )} \\
   = 2 \pi \, \delta( p^2/2 + U_p + E_B - \omega_X - \ell \omega_L ) \,.
\end{multline}
Here, the delta function ensures the energy conservation in this two-color interaction process and shows that the kinetic 
energy of the photoelectrons becomes discrete for sufficiently weak fields. In each of these \textit{sidebands} of 
the main photo line (that arise from the ionization by the XUV pulse), the modulus of the electron momenta is constant,
$|\vec p| =  p \to p_\ell = \sqrt{ 2 (\omega_X + \ell \omega_L - E_B - U_p ) }$, while these electrons may still exhibit an
(angular) distribution as function $\vartheta_p$ and $\varphi_p$. Using expression \eqref{eq:time:integral}, the 
transition amplitude for two-color ATI of an atom at position $\vec b$ by a XUV Bessel beam now becomes
\begin{align} \label{eq:amplitude:final1}
   \transamp_{\vec b}(\vec p)
   &= 2 \pi \sum_\ell \: \delta(p^2/2 + U_p + E_B - \omega_X  - \ell \omega_L) \: \transamp^{(\ell)} \, ,
\end{align}
and where
\begin{align}  \label{eq:amplitude:final2}
   \transamp^{(\ell)} 
   &= \scp{\vec p_\ell}{\phi_0} \: \int \! \twistint \, e^{i \vec k \cdot \vec b} \, \twistamp 
      \mathcal F_\ell(\vartheta_k,\varphi_k;\Lambda_X,\Lambda_L) 
\end{align}
{are} often referred to as \textit{partial amplitudes}. As seen from expression \eqref{eq:amplitude:final2}, the angular 
distribution of the photoelectrons is now determined by a convolution of the sideband amplitudes $\mathcal F_\ell$ 
with the Fourier coefficients of the vortex Bessel beam $\twistamp$ from Eq.~\eqref{eq:akm} and a phase factor that
just contains the impact vector $\vec b$. Indeed, the expressions \eqref{eq:amplitude:final1} and 
\eqref{eq:amplitude:final2} are one of our major results of this work, although they still describe the transition 
amplitude for a single atom at some (fixed) position $\vec b$ with regard to the beam axis.

\subsection{Photoionization probability of localized and macroscopically extended targets}
\label{sect:target}

We can use the two-color ATI amplitude \eqref{eq:amplitude:final1} to express the photoionization probability (per 
unit time) for an atom at position $\vec b$ within a vortex beam by
\begin{align} 
   \mathbb P_{\vec b}(\vec p) &= \frac{1}{T} \left| \transamp_{\vec b} (\vec p) \right|^2   \nonumber \\
   &= 2\pi \sum_\ell \delta(p^2/2+U_p+E_B-\omega_X - \ell \omega_L)
      \mathbb P^{(\ell)}_{\vec b} (\vec p) \, , 
   \label{eq:probability:b2}
\end{align}
if $T$ denotes here the interaction time of the atom with the two-color field, and if we make use of the usual 
interpretation of the delta function $\delta(0) = T/2\pi$ in the second line. Expression \eqref{eq:probability:b2} shows 
that the photoionization probability is a sum of partial probabilities
\begin{align}	\label{eq:probability:partial}
   \mathbb P^{(\ell)}_{\vec b}(\vec p) = \left|\transamp^{(\ell)} \right|^2
\end{align}
that describe the individual sidebands in the photoelectron spectrum. The partial probabilities 
{\eqref{eq:probability:partial}}
still refer, as before, to a single atom at impact vector $\vec b$ with regard to the 
beam axis. To further analyze and compare the photoelectron spectra and angular distribution with those obtained 
experimentally, we need to know (or assume) also the distribution of atoms in the overall cross section of the 
Bessel beam.

\subsubsection{Macroscopically extended targets}

If, for example, the twisted Bessel beam interacts with a homogeneous and (infinitely in the cross section of the beam) 
\textit{extended} target of atoms, we have to average the partial photoionization probabilities 
$\mathbb P^{(\ell)}_{\vec b}(\vec p)$ from Eq.~\eqref{eq:probability:partial} incoherently over all impact 
vectors $\vec b$,
\begin{align} \label{eq:def:probability:infinite}
   \mathbb P^{(\ell)} ( {\vec p} ) = \int \! \d^2 \vec b \;\: \mathbb P^{(\ell)}_{\vec{b}}(\vec p) \, ,
\end{align}
in order to obtain the partial photoionization probabilities, and the same is true for the total photoionization probability
{\eqref{eq:probability:b2}}.
Using Eqs.~\eqref{eq:amplitude:final2} and \eqref{eq:probability:partial}, we then obtain
\begin{widetext}
\begin{align}
   \mathbb P^{(\ell)} (\vec p) & =  |\scp{\vec p_\ell}{\phi_0}|^2
		\int \! \twistint \, \frac{\d^2 \vec k'_\perp}{(2\pi)^2} \, \d^2 \vec b \;\:
		a^*_{\varkappa m}(\vec k'_\perp) \: \twistamp\: e^{i \vec b\cdot (\vec k - \vec k')} \:
		\mathcal F_\ell^*(\vartheta_{k'},\varphi_{k'} {;\Lambda_X,\Lambda_L} ) 
		\: \mathcal F_\ell(\vartheta_k,\varphi_k {;\Lambda_X,\Lambda_L} ) \,.
\end{align}	
Since the impact vector occurs here only in the exponential, $e^{i \vec b \cdot (\vec k - \vec k') }$, the 
integration over $\vec b$ just gives rise to a delta function in momentum space, and the partial photoionization 
probability of sideband $\ell$ becomes
\begin{align}
   \mathbb P^{(\ell)}(\vec p)
   &=  |\scp{\vec p_\ell}{\phi_0}|^2 \,\int \frac{\d \varphi_k}{2\pi} \:
       |\mathcal F_\ell( \vartheta_k, \varphi_k; \Lambda_X , \Lambda_L ) |^2 \, .
\end{align}		
{Employing the expression for} the sideband amplitude \eqref{eq:sideband:tw}, we can now, furthermore, perform the integral over the azimuthal 
angle $\varphi_k$ and finally obtain for the partial photoionization probability
\begin{multline} \label{eq:probability:infinite}
   \mathbb P^{(\ell)}(\vec p)
   =  |\scp{\vec p_\ell}{\phi_0}|^2
      \left\{ \frac{p^2_\ell}{2} J_\ell^2
 	      \left[  \sin^2 \vartheta_p \left( 1 - 2 \sin^2 \frac{\vartheta_k}{2} + 2 \sin^4 \frac{\vartheta_k}{2} \right) 
 	              \:+\: \cos^2 \vartheta_p \sin^2 \vartheta_k		
	      \right] \right.  \\[0.15cm]
             \;-\; p_\ell  \sin \vartheta_p A_L J_\ell \:
	     \left[  J_{\ell + \Lambda_X \Lambda_L} \left( 1 - 2 \sin^2 \frac{\vartheta_k}{2} \right)
		     \:+\: (J_{\ell+1}+J_{\ell-1}) \sin^4 \frac{\vartheta_k}{2}
	     \right] \\[0.15cm]
   \left.  + \frac{A_L^2}{2}
             \left[  J_{\ell +\Lambda_X \Lambda_L}^2 \left( 1 - 2 \sin^2 \frac{\vartheta_k}{2} \right)
		     \:+\: (J_{\ell+1}^2+J_{\ell-1}^2) \sin^4 \frac{\vartheta_k}{2}
             \right] 
   \right\}  \; .
\end{multline}
\end{widetext}
Obviously, this probability depends on the cone opening angle $\vartheta_k$ of the (vortex) Bessel beam, while it is
independent of the topological charge $m$ for a macroscopically extended target.

\begin{figure}[!b]
\includegraphics[width=0.89\columnwidth]{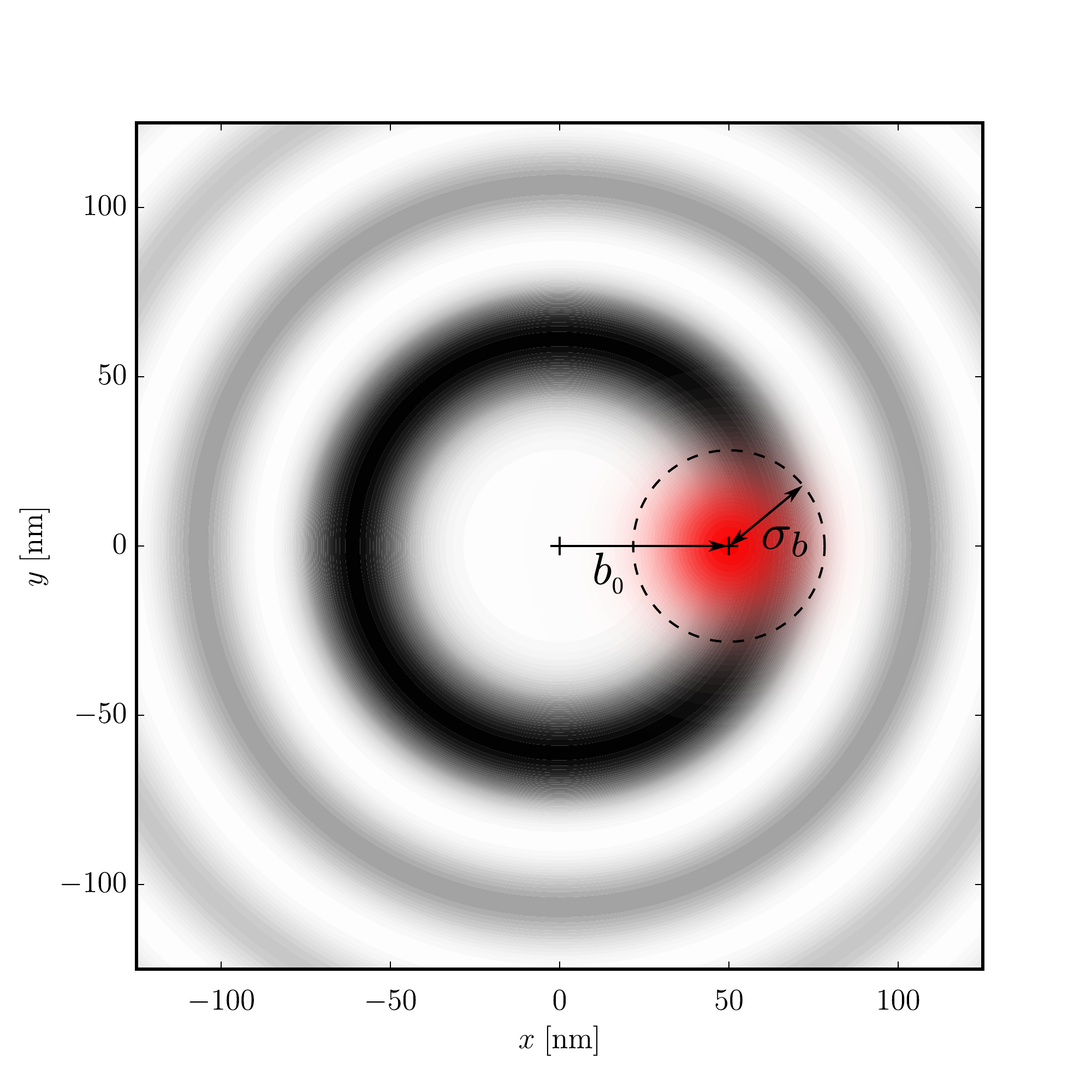}
\caption{Illustration of a localized atomic target of size  $\sigma_b$ that is displaced by the impact vector $b_0$ 
with regard to the center of the vortex beam.}
\label{fig:beam:target}
\end{figure}

\subsubsection{Localized targets}

Another (localized) target refers to a small cloud of atoms that is centered around the impact vector $\vec b_0$ 
in a plane perpendicular to the beam axis. We here assume a normalized Gaussian distribution of target atoms

\newpage
\begin{align}
   \rho(\vec b) =  \frac{1}{\sqrt{ 2\pi } \sigma_b } \exp \left\{ -\frac{(\vec b - \vec b_0)^2}{2 \sigma_b^2} \right\} \, 
\end{align}
where $\sigma_b$ denotes the (r.m.s.) size of the target, cf.~Fig.~\ref{fig:beam:target}. Without loss of generality,
moreover, we may assume that the impact vector $\vec b_0 = b_0 \vec e_x$ defines the $x$-axis and, hence, the angle 
$\varphi_p$ in the angular distribution of the photoelectrons below (and with the $z$-axis along the beam). For such a
localized target with distribution {$\rho(\vec b)$}, the partial photoionization probability becomes
\begin{align} \label{eq:probability:cloud}
   \mathbb P^{(\ell)}_\rho(\vec p) 
   &= \int \! {\d^2 \vec b} \, \rho(\vec b) \: \mathbb P_{\vec b}^{(\ell)}  (\vec p) \, ,
\end{align}
and where the integration over the target distribution below will be performed numerically.

\subsection{Dichroism in two-color fields}
\label{sect:dichroism}

In the previous section, we saw how the partial photoionization
{probabilities \eqref{eq:probability:partial}, \eqref{eq:probability:infinite} and \eqref{eq:probability:cloud}
describe} the 
yield of photoelectrons for a given sideband, as a function of the two emission angles $\vartheta_p$ and $\varphi_p$
{for different kinds of targets}.
Of course, these probabilities also depend on the spin- and orbital angular momenta of the incident XUV and assisting NIR
laser fields. To further understand how the coupling of these angular momenta affects the  relative photoionization 
probabilities, we may resort to different kinds of \textit{dichroism} signals as often used in describing the interaction 
of light with atoms, molecules and solids 
\cite{Kazansky:PRL2011,Mazza:NatCommun2014,Lux:AngewChemie2012}.

\subsubsection{Circular dichroism for plane-waves}

Let us start from the (atomic) \textit{circular} dichroism which has been frequently used in characterizing the 
photoelectron emission if both, the XUV and the assisting NIR fields are described by plane waves. For two 
plane waves, as shown in Sect.~\ref{subsubsect:theory:plane}, the two-color ATI amplitude \eqref{eq:amplitude:plane} 
and, hence, the corresponding photoionization probability only depends on the product of the two spin angular momentum (SAM) projections, 
i.e.~the helicities of the XUV and laser photons. While there are four possible combinations of these helicities, 
only two $\Lambda_X\Lambda_L \:=\: \pm1$ are distinguishable from each other. For the interaction of atoms with 
two plane waves, we can therefore define just \textit{one} dichroism signal, 
\begin{align} \label{eq:def:CD}
   \mathrm{CD} 
   = \frac{      \mathbb P({\vec p};\Lambda_X ,  \Lambda_L) 
            \;-\; \mathbb P (\vec p; \Lambda_X , -\Lambda_L ) 
	  }{      \mathbb P(\vec p; \Lambda_X,\Lambda_L ) \:+\: \mathbb P (\vec p;  \Lambda_X, -\Lambda_L )} \,,
\end{align}
commonly known also as circular dichroism \cite{Kazansky:PRA2012}, and which is a function of the photoelectron emission 
angles $\vartheta_p$ and  $\varphi_p$, respectively. This circular dichroism can be defined uniquely for each sideband as
long as the incident light beams are sufficiently monochromatic.

\begin{table*}[t]
\caption{Definition of the seven dichroism signals for the two-color ionization of atoms by a vortex (Bessel) beam and 
a plane-wave laser field.}
\label{tab:1}
\begin{tabular}{p{9.6cm}cr ll}                    \\[-0.25cm]
\toprule[1.5pt]                                   \\[-0.25cm]
  Dichroism due to a flip of \ldots        
    &\hspace*{0.5cm} & Definition \\[0.1cm]  \midrule[1pt]   \\[-0.25cm]
  the helicity of the assisting NIR laser field.  
    && $\displaystyle \dichroism{\Lambda_L}$ 
    &$=$&  {\large $\frac{\mathbb P(\vec p, m, \Lambda_X,\Lambda_L) - \mathbb P(\vec p, m, \Lambda_X,-\Lambda_L)  
                        }{\mathbb P(\vec p, m, \Lambda_X,\Lambda_L) + \mathbb P(\vec p, m, \Lambda_X,-\Lambda_L) }$ 
	   } \\[0.3cm]
  the helicity of the XUV photons. 
    && $\displaystyle \dichroism{\Lambda_X}$ 
    &$=$&  {\large $ \frac{\mathbb P(\vec p, m, \Lambda_X,\Lambda_L) - \mathbb P(\vec p, m, -\Lambda_X,\Lambda_L)  
                         }{\mathbb P(\vec p, m, \Lambda_X,\Lambda_L) + \mathbb P(\vec p, m, -\Lambda_X,\Lambda_L) }$  
	   } \\[0.3cm]
  the projection of the orbital angular momentum. 
    && $\displaystyle \dichroism{m}$ 
    &$=$&  {\large $ \frac{\mathbb P(\vec p, m, \Lambda_X,\Lambda_L) - \mathbb P(\vec p, -m, \Lambda_X,\Lambda_L)  
                         }{\mathbb P(\vec p, m, \Lambda_X,\Lambda_L) + \mathbb P(\vec p, -m, \Lambda_X,\Lambda_L) }$   
	   } \\[0.3cm]
  the helicity and the orbital angular momentum of the XUV Bessel beam. This is equivalent to just a flip of the 
  projection of the total angular momentum. 
    && $\displaystyle \dichroism{m\Lambda_X}$ 
    &$=$&  {\large $ \frac{\mathbb P(\vec p, m, \Lambda_X,\Lambda_L) - \mathbb P(\vec p,- m, - \Lambda_X,\Lambda_L)  
                         }{\mathbb P(\vec p, m, \Lambda_X,\Lambda_L) + \mathbb P(\vec p,- m, - \Lambda_X,\Lambda_L) }$ 
	   } \\[0.3cm]
  the helicities of both the laser and XUV photons. For two plane waves this dichroism signal is always 
  zero because of the symmetry. 
    && $\displaystyle \dichroism{\Lambda_X\Lambda_L}$ 
    &$=$&  {\large $ \frac{\mathbb P(\vec p, m, \Lambda_X,\Lambda_L) - \mathbb P(\vec p, m, - \Lambda_X,-\Lambda_L)  
                         }{\mathbb P(\vec p, m, \Lambda_X,\Lambda_L) + \mathbb P(\vec p, m, - \Lambda_X,-\Lambda_L) }$ 
	   } \\[0.3cm]
  the projection of the orbital angular momentum of the Bessel beam and of the helicity of the laser field. 
    && $\displaystyle \dichroism{m\Lambda_L}$ 
    &$=$&  {\large $ \frac{\mathbb P(\vec p, m, \Lambda_X,\Lambda_L) - \mathbb P(\vec p,- m,  \Lambda_X,-\Lambda_L)  
                         }{\mathbb P(\vec p, m, \Lambda_X,\Lambda_L) + \mathbb P(\vec p,- m,  \Lambda_X,-\Lambda_L) }$
	   } \\[0.3cm]
  all three projections of the angular momenta simultaneously.
    && $\displaystyle \dichroism{m\Lambda_X\Lambda_L}$
    &$=$&  {\large $ \frac{\mathbb P(\vec p, m, \Lambda_X,\Lambda_L) - \mathbb P(\vec p,- m,  -\Lambda_X,-\Lambda_L)  
                         }{\mathbb P(\vec p, m, \Lambda_X,\Lambda_L) + \mathbb P(\vec p,- m,  -\Lambda_X,-\Lambda_L) } $ 
	   } \\[0.3cm]
  \bottomrule[1.5pt]
\end{tabular}
\end{table*}

\subsubsection{Dichroism signals for (vortex) Bessel beams}	

For vortex Bessel beams, the photoionization probability depends not only on the SAM of the XUV ($\Lambda_X=\pm1$)
and laser beams ($\Lambda_L=\pm1$) but also on the \textit{orbital} angular momentum $\pm m$ of the XUV photons. With three
angular momenta, we can form eight combinations of $(m,\Lambda_X,\Lambda_L)$ by just changing the sign of one 
or more of these quantum numbers. This enables us to define \textit{seven} different dichroism signals for the two-color
ionization of atoms by a vortex and a plane-wave beam since one of the combinations, 
$\mathbb P (\vec p,|m|,\Lambda_X=+1,\Lambda_L=+1)$, should occur as reference. For example, the dichroism that is 
associated with a flip of the projection of the orbital angular momentum $\pm m$ is easily defined by
\begin{align}
   \dichroism{m} 
   &= \frac{\mathbb P(\vec p,m,\Lambda_X,\Lambda_L)  \:-\: \mathbb P (\vec p,-m, \Lambda_X,\Lambda_L)
          }{\mathbb P (\vec p,m,\Lambda_X,\Lambda_L) \:+\:  \mathbb P(\vec p,-m ,\Lambda_X,\Lambda_L)}  \,.
\end{align}
Very similarly, we can define all the other dichroism signals as associated with some flip in the helicity and/or OAM 
quantum numbers, and which are displayed explicitly in Table~\ref{tab:1}~. As for the circular dichroism, all these 
(seven) dichroism signals generally depend for a localized target on the photoelectron emission angles $\vartheta_p$ and 
$\varphi_p$ as well as on the particular sideband $\ell$.

For sufficiently extended targets ($\sigma_b \varkappa \gg 1$), however, the photoionization probability and, hence,	
the dichroism signals above become independent of the (projection of the) orbital angular momentum or topological charge,
$m$. This can be seen for instance from the analytical expression for the photoionization probability for infinitely 
extended targets \eqref{eq:probability:infinite}, which is independent of $m$. For large targets, therefore, all the 
dichroism signals will depend just on the product $\Lambda_X\:\Lambda_L$ of the two helicities and, thus, all signals 
with $\Lambda_X\:\Lambda_L\:=\: +1$ must vanish in this limit, 
$\dichroism{\Lambda_X\Lambda_L} = \dichroism{m \Lambda_X \Lambda_L} = \dichroism{m} =0$. Moreover, all other signals with
$\Lambda_X\:\Lambda_L\:=\: -1$ then coincide with the usual circular dichroism,
$\dichroism{\Lambda_L} = \dichroism{\Lambda_X} = \dichroism{m \Lambda_X} = \dichroism{m\Lambda_L} $, 
cf.~Eq.~\eqref{eq:def:CD}. For extended targets, a nonzero dichroism signal can be observed only if just \textit{one} 
of the helicities $\Lambda_X$ or $\Lambda_L$ is changed.

\begin{figure*}[!t]
\begin{minipage}{0.49\textwidth}
\includegraphics[width=0.99\columnwidth]{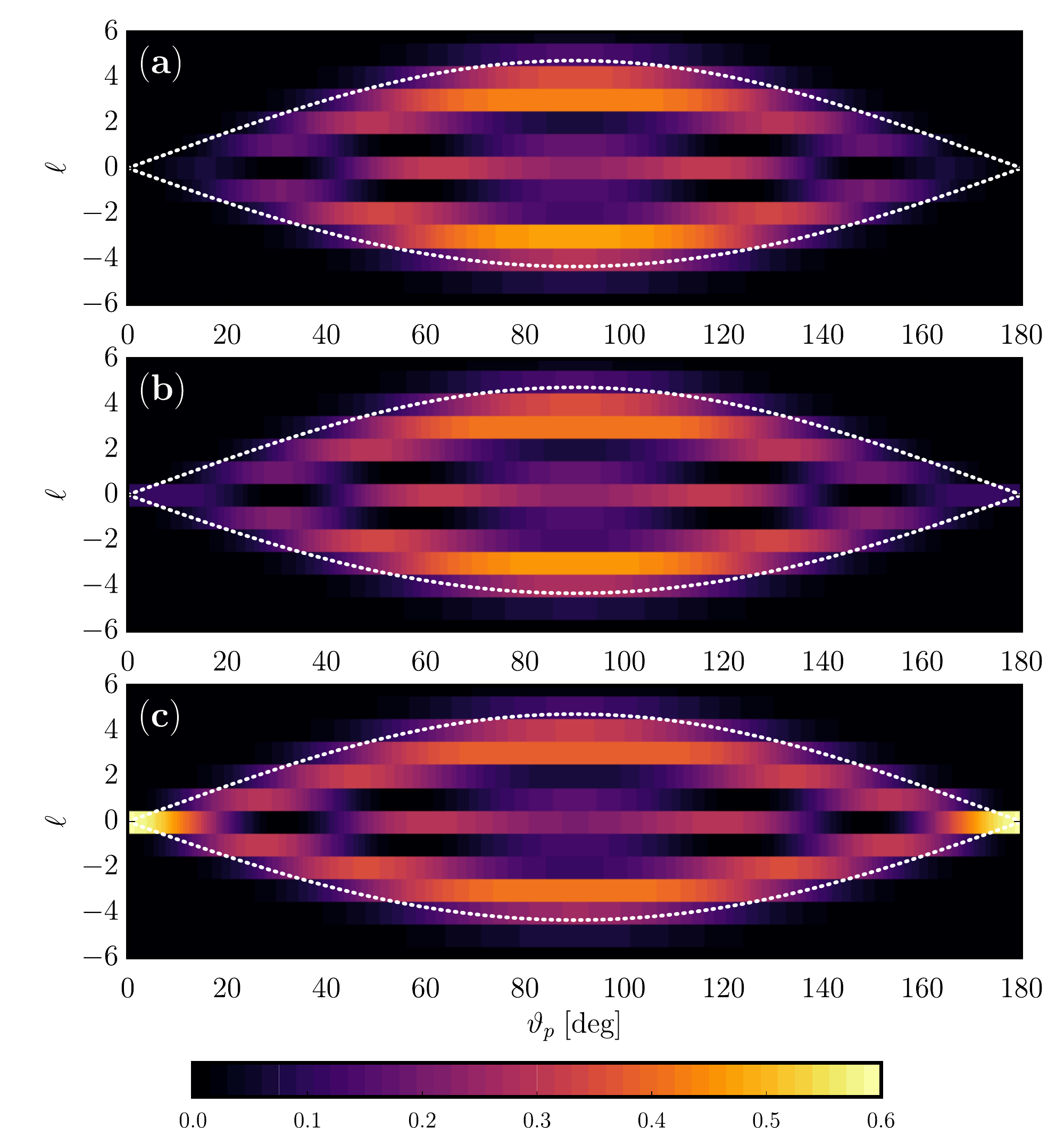}
\end{minipage}
\begin{minipage}{0.49\textwidth}
\caption{(Color online). Contourplot of the two-color ATI photoionization probability $\mathbb P^{(\ell)}(\vartheta_p)$
as function of the emission angle $\vartheta_p$ and the sideband number $\ell$, i.e.~the net number of absorbed or emitted
laser photons by the outgoing photoelectrons. Results, encoded by colors (gray scale), are given for an infinitely 
extended target and are shown for (a) a plane-wave XUV beam as well as (b,c) a XUV Bessel beam with cone opening 
angles $\vartheta_k=0.2$ rad and $\vartheta_k=0.5$ rad, respectively. All these probabilities are independent of 
the topological charge $m$ of the Bessel beams because of the target size. The calculations were performed for the XUV 
frequency $\omega_X = 3\,{\rm a.u.} = \unit{81.6}{\electronvolt}$ as well as for an assisting NIR laser field with
frequency $\omega_L = {0.05}\,{\rm a.u.} = \unit{1.36}{\electronvolt}$ and amplitude $A_L=0.1$ of the vector 
potential.}
\label{fig:probability}
\end{minipage}
\end{figure*}

\begin{figure*}
\begin{center}
\includegraphics[width=0.95\textwidth]{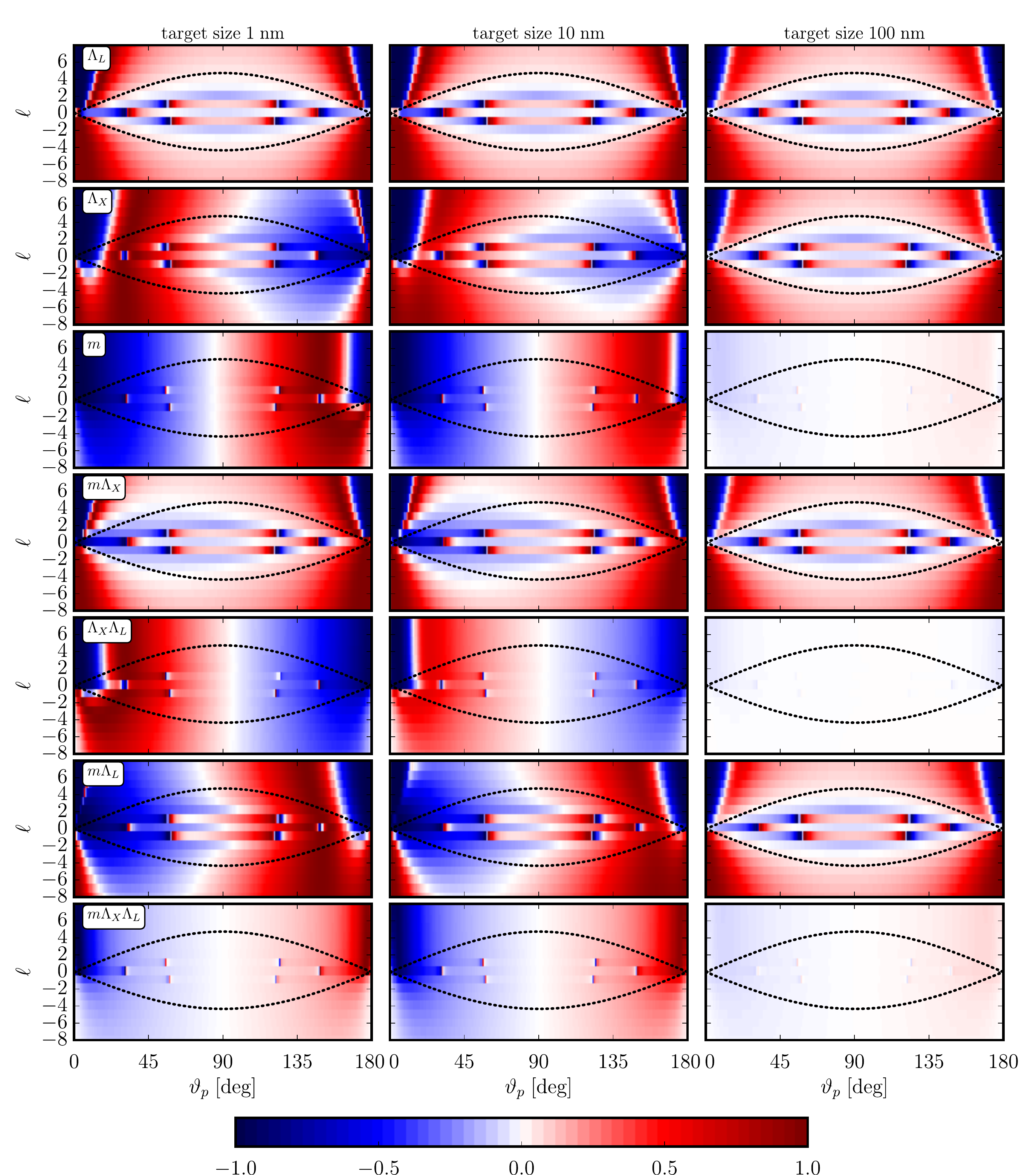}
\end{center}
\caption{(Color online). Contourplot of the dichroism as function of the emission angle $\vartheta_p$ and the 
sideband number $\ell$ of the emitted electrons. Results are shown for the seven dichroism signals from Table~~\ref{tab:1}
(rows 1-7), encoded by colors, and for the three target sizes $\sigma_b=\unit{1}{\nano\metre}$ 
(left column), $\sigma_b=\unit{10}{\nano\metre}$ (middle column) and $\sigma_b=\unit{100}{\nano\metre}$ (right column), 
respectively. The photoelectrons are observed at the azimuthal angle $\varphi_p=\unit{90}{\degree}$ with respect to the 
impact vector $b_0=\unit{30}{\nano\metre}$ as center of the target. All calculations were performed for a Bessel beam 
with photon energy $\omega_X = {3}\,{\rm a.u.} = \unit{81.6}{\electronvolt}$, opening angle $\vartheta_k=0.2$ and 
orbital angular momentum projection $m=4$. The black dotted curves indicate the cut-off values of the number of 
sidebands as given analytically by Eq.~\eqref{eq:cutoffs}. }
\label{fig:polar:dichroism}
\end{figure*}

\begin{figure*}
\begin{center}
\includegraphics[width=0.95\textwidth]{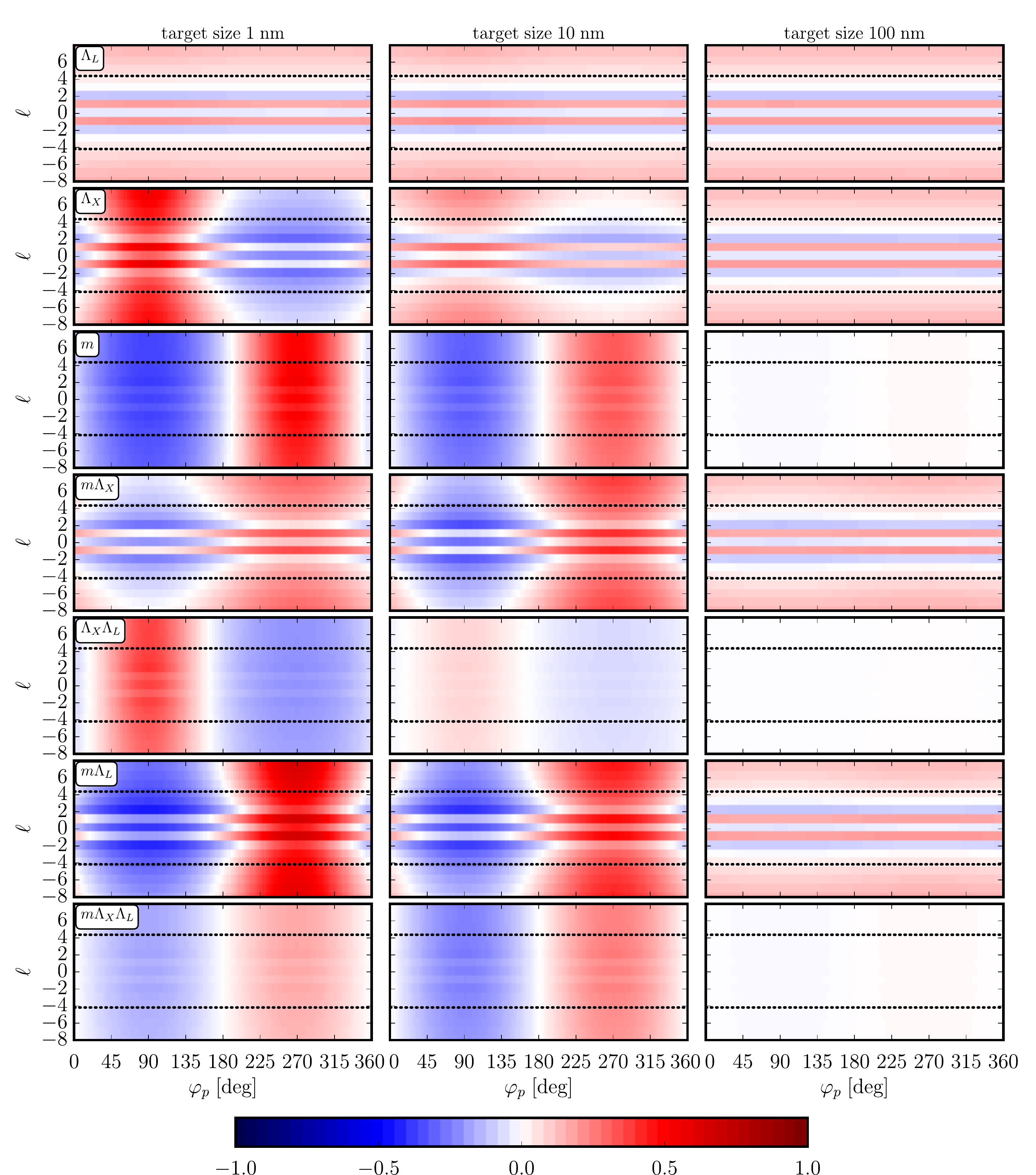}
\end{center}
\caption{(Color online). The same as figure~\ref{fig:polar:dichroism} but as function of the azimuthal angle 
$\varphi_p$ and for a Bessel beams with photon energy $\omega_X = {5}\,{\rm a.u.} = \unit{136}{\electronvolt}$.
Here, the photoelectrons are observed under the polar angle $\vartheta_p=\unit{45}{\degree}$ with regard to the
beam axis and for a impact parameter $b_0 = \unit{25}{\nano\metre}$.}
\label{fig:azimuthal:dichroism}
\end{figure*}

\section{Results and Discussion}
\label{sect:results}

In the last section, we have analyzed the transition amplitude and photoionization probability for the two-color ATI of 
atoms by a vortex (Bessel) beam and combined with an intense plane-wave (NIR) laser field. Emphasize was placed here
on the evaluation of this amplitude and the sideband structure of the {central photoline} due to the interaction of 
the emitted electrons with the NIR field. We also introduced various dichroism signals by flipping the projections of 
the spin and orbital angular momenta of the involved fields in order to quantify the dependence of the photoionization 
probability upon the angular momentum properties of the incident light beams.

As discussed above, the two-color ATI probability crucially depends for Bessel beams also on the size of the atomic target.
To better understand the influence of this target size, detailed computations were performed for the photoionization 
of the $4s$ valence electron of Ca$^+$ ions with binding energy $E_B = \unit{11.7}{\electronvolt} = {0.43}\,{\rm a.u.}$. 
Simple core-Hartree wave functions in a screened potential have been applied to calculate all the necessary (one-electron) atomic matrix elements \cite{Fritzsche:CPC2012}, cf.~\eqref{eq:atomic}.

{In Fig.~\ref{fig:probability}, we display the two-color ATI probability $\mathbb P^{(\ell)}(\vartheta_p)$
as function of the emission angle $\vartheta_p$ (horizontal axis) and the sideband number $\ell$ (vertical axis). The
sideband $\ell$ gives directly the net number of laser photons from the NIR field that are either absorbed or emitted
by the outgoing photoelectrons. The ATI probabilities are encoded by colors and are shown for an infinitely 
extended target. Results are compared for the two-color ATI by a plane-wave XUV beam (upper panel) as well as for a
XUV Bessel beam with cone opening angles $\vartheta_k=0.2$ rad (middle panel) and $\vartheta_k=0.5$ rad (lower panel), 
respectively. In these computations, we applied a XUV beam with the rather high frequency
$\omega_X = 3\,{\rm a.u.} = \unit{81.6}{\electronvolt}$ and for a NIR laser field with 
$\omega_L = {0.05}\,{\rm a.u.} = \unit{1.36}{\electronvolt}$ and amplitude $A_L=0.1$. 
As seen from Fig.~\ref{fig:probability}, the photoelectron distributions {exhibit} an almond-like shape for which the
largest number of sidebands occurs at $\vartheta_p \,\simeq\, \unit{90}{\degree}$, while only the main photoline 
($\ell=0$) is seen along the beam axis, i.e.~for $\vartheta=\unit{0}{\degree}$ and $\unit{180}{\degree}$. This shape of the
photoelectron distributions is well predicted also by the semiclassical cutoffs, Eq.~\eqref{eq:cutoffs}, as indicated by 
the white dotted curves in the figures.}

{For the two-color ATI by a plane-wave XUV beam, the calculated photoelectron distribution agrees qualitatively well 
with the calculations by Kazansky and coworkers \cite{Kazansky:PRA2012}. While no photoelectrons are seen in this case along the axis for plane-waves [cf.~Fig.~\ref{fig:probability} a],
this changes in the case of a twisted Bessel XUV 
beams in Figs.~\ref{fig:probability} b,c. For such Bessel beams, the photoionization probabilities along the beam axis
increases with the cone opening angle $\vartheta_k$. We note that the plane-wave result is of course recovered in the 
paraxial approximation for $\vartheta_k\to 0$.}

{To analyze the localization effects of the target, we use the different dichroism signals as defined in Section 
\ref{sect:dichroism} and Table \ref{tab:1}. Figure~\ref{fig:polar:dichroism}, for example, shows these dichroism signals 
as function of the emission angle $\vartheta_p$ and sideband number $\ell$ of the emitted electrons, and with the 
magnitude of the signals encoded by colors in the (seven) rows of the figure. We here applied a Bessel beam of the same 
frequency $\omega_X = \unit{81.6}{\electronvolt}$ and opening angle $\vartheta_k=0.2$ as in 
Fig.~\ref{fig:probability}, and with the projection of the angular momentum $m=4$. 
Detailed computations are carried out for the three target sizes $\sigma_b=\unit{1}{\nano\metre}$ (left column), 
$\sigma_b=\unit{10}{\nano\metre}$ (middle column) and $\sigma_b=\unit{100}{\nano\metre}$ (right column),
and for photoelectrons that are observed at the azimuthal angle $\varphi_p=\unit{90}{\degree}$ with regard to the 
impact vector $\vec b_0 = b_0 \, \vec e_x$ with $b_0=\unit{30}{\nano\metre}$ as the center of the target.
While all the dichroism signals are quite different 
from each other for a small target (left column) and, hence, sensitive to the particular localization of the target, these
differences become less pronounced as the target size increases. For target sizes (much) larger than the typical width 
of the rings in the Bessel beam, moreover, the dichroism signals approach the two limits: They either vanish identically
if the product of the helicities of the two-color field is positive, $\Lambda_X\:\Lambda_L\:=\: +1$ 
{[cf.~the right panels of rows 3, 5 and 7]}, or these signals coincide with the known circular dichroism for 
$\Lambda_X\:\Lambda_L\:=\: -1$ {[cf.~the right panels of rows 1, 2, 4, and 6]}.
Let us note also that the (usual) 
circular dichroism signal in row~1 appears to be rather insensitive to the size of the target. In fact, these dichroism
signals do not depend much on the details of the applied matrix elements as, in the electric-dipole approximation,
the prefactor in Eq.~\eqref{eq:amplitude:final2} cancel in the ratio that is formed by any dichroism. 
Finally, the black dotted curves indicate the cut-off values of the number of sidebands as given analytically 
by Eq.~\eqref{eq:cutoffs}.}
{That means, while we can calculate a dichroism signal also outside the almond-shaped
area, its measurement might be challenging since the photoionization probability
is very small in these regions, cf.~Fig.~\ref{fig:probability}.}

{Due to the phase of the XUV Bessel beam, a localization of the target affects not only the (polar) angular emission
of the photoelectrons but may results also in a non-trivial azimuthal distribution. Therefore,
Fig.~\ref{fig:azimuthal:dichroism} shows the same as Fig.~\ref{fig:polar:dichroism} but here as function of the 
azimuthal angle $\varphi_p$ and for a Bessel beams with slightly higher photon energy 
$\omega_X = {5}\,{\rm a.u.} = \unit{136}{\electronvolt}$ and for a target centered at $b_0 = \unit{25}{\nano\metre}$.
In this figure, the photoelectrons are assumed to be observed
under the polar angle $\vartheta_p=\unit{45}{\degree}$ with regard to the beam axis.
An azimuthal anisotropy of the ATI probabilities is found for the localized targets as it was obtained before for 
the azimuthal distribution of photoelectrons \cite{Matula:JPB2013}. This anisotropy of the ionization probabilities 
occurs of course also in the dichroism signals, while no azimuthal dependence appears for the usual circular dichroism 
(first row). As for the polar-angle dependence in Fig.~\ref{fig:polar:dichroism}, all dichroism signals become either
zero or simply approach the circular dichroism for sufficiently large targets. }

\section{Summary and Conclusions}
\label{sect:summary}

{In this work, we investigated the two-color ATI of atoms and ions if light from a weak XUV Bessel beam is combined 
with a strong NIR laser field. While the emission of photoelectron occurs due to the weak XUV beam, the energy and 
angular distribution of the photoelectrons is affected by the plane-wave NIR field due to a net absorption or emission of
one or several laser photons. Thus, the interaction of the atoms with such a two-color field results
in sidebands to the
normal photoline which are affected not only by the intensity and temporal structure of the NIR field but also by
the  location and extent of the atomic target as well as by the spin and orbital angular momenta of the two fields involved.}

{Emphasis in our analysis has been placed upon the energy spectra and angular emission of the photoelectrons as well as on the asymmetry in the photoelectron spectra if some of the SAM or OAM components of the fields are flipped relative to each
other. For a XUV Bessel beam and a plane-wave NIR field, seven different dichroism signals can be defined. These signals
differ for localized target but become either zero or coincide with the usual circular dichroism for macroscopically 
extended targets, similar as for Bessel beams with small opening angle.}
Our investigation of two-color strong field ATI with XUV vortex Bessel beams and 
the discussion of the seven different dichroism signals
opens up avenues for future investigations of the interaction of
atomic and molecular targets
with twisted light in the high-intensity regime.

\section{Acknowledgments}

This work was supported by the DFG priority programme 1840, "Quantum Dynamics in Tailored Intense Fields".

%


\end{document}